\documentclass[a4paper,british,english,oldversion]{aa}
\usepackage[T1]{fontenc}
\usepackage[latin9]{inputenc}
\usepackage{color}
\usepackage{amsmath}
\usepackage{amssymb}
\usepackage{graphicx}
\usepackage[authoryear]{natbib}

\makeatletter

\pdfpageheight\paperheight
\pdfpagewidth\paperwidth

\providecommand{\tabularnewline}{\\}
\newcommand{\lyxdot}{.}

\usepackage{xcolor}
\usepackage{ifthen}
\authorrunning{T. Dermine et al.}
\titlerunning{Post-AGB Stars with Circumbinary Disks}
 \date{}

\usepackage{amsbsy} 
\usepackage{setspace}
\definecolor{Mygrey}{gray}{0.5}

\makeatother

\usepackage{babel}
\begin{document}

\title{Post-AGB Stars with Circumbinary Discs}

\author{Tyl Dermine\inst{1,2}\and Robert G. Izzard\inst{1}\and Alain
Jorissen\inst{2}\and Hans Van Winckel\inst{3}}

\institute{Argelander-institut f\"ur Astronomie, University of Bonn, Auf dem
Hügel 71, D-53121 Bonn, Germany\and  Institut d'Astronomie et d'Astrophysique,
Université Libre de Bruxelles, Belgium\and  Instituut voor Sterrenkunde,
K.U.Leuven, Belgium}

\abstract{Circumbinary discs are commonly observed around post-asymptotic giant
branch (AGB) systems and are known to play an important role in their
evolution. Several studies have pointed out that a circumbinary disc
interacts through resonances with the central binary and leads to
angular momentum transfer from the central binary orbit to the disc.
This interaction may be responsible for a substantial increase in
the binary eccentricity. We investigate whether this disc eccentricity-pumping
mechanism can be responsible for the high eccentricities commonly
found in post-AGB binary systems. }

\maketitle

\section{Introduction}

Post-AGB stars just left the AGB phase and evolve rapidly to hotter
effective temperatures at constant luminosity, but they are not yet
hot enough to ionise circumstellar material ejected during the AGB
phase. Some post-AGB stars also show a near-infrared emission due
to thermal radiation of hot circumstellar dust. The colour temperature
of the dust is a good indication that there must by circumstellar
dust close to the star and this SED characteristic is always related
to binarity \citep[and references therein]{VanWinckel-2009}. The
extremely narrow CO emission lines (a few km s$^{-1}$ width, e.g.
\citealp{Jura-1995}), the presence of large (sub-micron) grains \citep{Gielen-2011}
and the detection of Oxygen-rich crystalline silicates \citep{Waters-1998a}
are distinct characteristics best explained by the long-term processing
of dust in a stable circumbinary (CB) disc.

However, a key problem still remains regarding the inability of population-synthesis
models to reproduce the observed large eccentricities of post-AGB
stars, similarly observed in barium stars \citep{Pols-03,Izzard-2010}.
In the previous evolutionary phase (i.e. the AGB), the very large
stellar radii lead to efficient circularisation of the binary orbit
by tides \citep{Zahn-1977}. Systems too close to accommodate an AGB
star undergo Roche-lobe-overflow (RLOF) mass transfer, in most cases
in an unstable regime which leads the system into a common envelope
(CE). Due to friction between the stellar cores and the CE, the envelope
may be ejected, and the orbit shrunk and circularised. Post-AGB binaries
with orbital periods shorter than $1,000$~d are expected to be circular
whereas observations reveal instead systems with periods from about
$10^{2}$~d to $3,000$~d, often in eccentric orbits. A mechanism
that increases the binary eccentricity is therefore required.

If tides are as efficient as predicted \citep{Zahn-1977}, a mechanism
to increase the eccentricity must take place during the AGB or post-AGB
evolution. Different mechanisms have been investigated such as enhanced
mass-transfer at periastron \citep[i.e. differential mass loss during an orbital period;][]{Soker-2000},
eccentricity pumping induced by a wind-RLOF hybrid mass transfer \citep{Bonacic-2008}
and a kick to the newly-born white dwarf \citep{Izzard-2010}.

In this paper we investigate the possibility that CB discs cause the
large eccentricities observed in some post-AGB systems. The current
knowledge of CB discs is summarised in Section~\ref{sec:Circumbinary disk}.
In Section~\ref{sec:Modelling post-AGB with CB discs} we present
our results and compare them to observed post-AGB periods and eccentricities.
Section~\ref{sec:Discussion} discusses successes and potential problems
of our model. Conclusions are drawn in Section~\ref{sec:Conclusions}.

\section{Circumbinary discs}

\label{sec:Circumbinary disk}

In this section we enumerate the classes of stars known or suspected
to be binaries that possess dust discs.

\subsection{Observed disc properties}

\textit{\textcolor{red}{\label{sub:Observed Disc Properties}}}Dust
discs are observed in very different classes of stars known to be
binaries \citep{Chesneau-2011}. Stable discs are observed around
most known post-AGB binaries \citep{VanWinckel-2003,DeRuyter-2006,VanWinckel-2006}.
They are increasingly being identified around the central stars of
planetary nebulae, e.g. NGC~2346 \citep{Costero-1986}. Other evidence
exists such as eclipses from dust discs (in NGC~2346, CPD-56$^{\circ}$8032
and M2-29, \citealt{Hajduk-2008,Gesicki-2010}). Some symbiotic stars
\citep{Angeloni-2007}, hydrogen-deficient binary \citep{Netolicky-2009}
and young stellar objects \citep{Deroo-2007a} are also known to possess
long-lived CB discs. Around some silicate J-type stars, the disc is
resolved \citep{Deroo-2007}. Some AGB stars (e.g. X Her ; \citealp{Kahane-1996})
have a very slowly expanding wind (a few km s$^{-1}$), however none
is a confirmed binary system. Keplerian discs are also observed in
B{[}e{]} stars \citep{Meilland-2007}; but only few B{[}e{]} stars
are confirmed binaries and it remains unclear whether the B{[}e{]}
behavior is related to binarity \citep{Lamers-1998}. Finally, discs
have been discovered around white dwarfs \citep[e.g.][]{Gaensicke-2006}.
Surveys for such discs around the oldest white dwarfs have been unsuccessful
\citep{Kilic-2009}, but around 15\% of local white dwarfs show metal-rich
material in their photosphere, indicative of accretion from a residual
dust reservoir \citep{Sion-2009}, although this reservoir may not
necessarily originate from binary interaction.

The variety of classes of objects that host CB discs and the wide
range of orbital periods of the central stars (from 115.9~d for SAO~173329
- \citealp{VanWinckel-2000:a} - to 2597~d for U~Mon; \citealp{Pollard-1995})
show that they are very common around binary systems and easily formed
\emph{in late-type mass-transfer} systems. The structure of discs
surrounding young stellar objects and post-AGB stars is somewhat similar
\citep{DeRuyter-2006}, which is rather remarkable because of their
(possibly) very different formation mechanism.

The stable nature of the dust disc in post-AGB systems can be inferred
from the observation that none of them shows evidence for a current
dusty mass-loss while the dust excess starts near dust-sublimation
radius; the near-IR emitting material must be gravitationally bound
in the system. In addition, post-AGB systems show depletion patterns,
i.e. the gas in the disc is separated from the dust and subsequently
reaccreted onto the star \citep{Waters-92,Maas-2005}. This re-accretion
process occurs more efficiently if circumstellar gas and dust are
trapped in a stable disc \citep{VanWinckel-2006,DeRuyter-2006,Gielen-2008}.
Furthermore the presence of \emph{near}-IR (from about $5$ to $20\,\mu$m)
excesses indicates that in all systems the circumstellar shell is
not freely expanding but stored in a disc. The detection of cool (from
about $100$ to $150\,$K) Oxygen-rich \emph{crystalline} silicate
dust grains \citep{Waters-1998a} as well as the presence of large
(sub-micron) grains are other indications of the long-lived nature
of these discs \citep[e.g.][]{Angeloni-2007,Gielen-2008}. In some
cases, Keplerian rotation is detected at least in the inner disc region
\citep{Bujarrabal-2005}.

\subsection{Formation}

\label{sub:Formation of the CB disc}The formation of CB discs and
their impact on the central star is not well understood. Most likely,
CB discs in post-AGB stars remain from CE ejection or form during
mass-transfer events in a binary system. The former case is supported
by population synthesis models which show that most post-AGB systems
that possess a CB disc are post CE systems. The possibility to form
a CB disc from a fraction of the ejected CE has already been suggested
from detailed CE models \citep{Sandquist-1998,Nordhaus-2006,Kashi-2011}.
It turns out that the CE is not necessarily completely ejected, so
some gas remains around the system. While CE evolution is expected
to lead systems to orbital periods of hours to hundreds of days, no
disc has been observed around systems with periods shorter than a
hundred days. 

On the other hand, many studies have investigated the possibility
that matter flows through the second Lagrangian point L$_{2}$ during
RLOF \citep[e.g.][]{Podsiadlowski-1992,Frankowski-2007a,VanRensbergen-2011},
possibly because of radiation pressure which could reshape the Roche
equipotentials by a reduction of the effective gravity of the mass-losing
star \citep{Schuerman-1972,Frankowski-2001,Dermine-2009}. Hydrodynamical
simulations have shown that this can lead to the formation of a CB
disc in a close binary system \citep{Sytov-2009}. However, the disc
may not be stable and may disappear as soon as the mass transfer stops.
Two-dimensional hydrodynamical simulations show that, for a typical
slow and massive wind from an AGB star, the flow pattern is similar
to Roche lobe overflow, and that a small fraction of the mass transferred
to the companion flows through L$_{2}$ \citep{deVal-borro-2009,Mohamed-2010}.
The companion focuses the ejected material into the equatorial plane
of the system \citep{Theuns-Jorissen-93,Theuns-1996,Mastrodemos-1998}.
However, the possibility of forming a stable CB disc from a mass-losing
giant star has not yet been confirmed by simulations.

\subsection{Evolution}

\label{sub:Evolution}Surveys performed at wavelengths as long as
100~$\mu$m show that none of the post-AGB systems with a disc is
observed to be surrounded by an outflow remaining from the AGB phase
\citep{Gielen-2011}. This means that the observed post-AGBs have
left the AGB phase long enough ago for the last ejecta of the AGB
star to reach a distance at which the emission at 100~$\mu$m cannot
be detected. A lower limit of the disc lifetime can then be inferred.
Using the code DUSTY \citep{Ivezic-1997} for a post-AGB star with
an effective temperature of $4,000\,$K, a luminosity of $5,000\,$L$_{\odot}$
and a wind composed only of silicate grains, we deduce a distance
of about 0.4~pc. Assuming a typical wind velocity of $15$~km~s$^{-1}$,
this timescale is approximately $2.5\times10^{4}\,$yr.

It is known that the CB-disc mass decreases during post-AGB and PN
phases because material flows away from the outer part of the disc
and reaccretes on the central stars. \citet{Gesicki-2010} pointed
out the sharp decrease in CB-disc mass as the star ages. Finally,
the discs observed around cooling white dwarfs have very low masses
because all the gas was probably expelled and only the dust remains.
An upper limit of the disc lifetime can then be estimated from the
time since the mass transfer stopped, i.e. the end of the AGB phase,
given by the timescale to cross the post-AGB and PN phases. While
it has been estimated for single stars (of the order of $1,000-10,000$~yr;
\citealp{Bloecker-1995}), this timescale is very uncertain for binaries.
In single stars, the post-AGB phase ends as soon as the thin envelope
remaining from the AGB phase is entirely ejected. However if re-accretion
occurs, e.g. from a CB disc, the envelope ejection could take much
longer, possibly up to $10^{5}$~yr. From the observed strongly depleted
post-AGB stars in the LMC \citep{Reyniers-2007,Gielen-2009}, we can
infer a dilution of their envelopes by a factor of about 10. As the
typical envelope mass is about $10^{-3}\,$M$_{\odot}$ \citep{Bloecker-1995},
$10^{-2}\,$M$_{\odot}$ must have been accreted from a CB disc. With
a typical mass-loss rate of $10^{-7}\,$M$_{\odot}$yr$^{-1}$, the
accreted mass therefore extends the post-AGB lifetime by about $10^{5}\,$yr.

The disc lifetime is determined by two competing mechanisms: (i) disc
accretion \citep{Hartmann-1998}, and (ii) photoevaporation of the
surface layer of the gas disc due to far- and extreme-ultraviolet
radiation from the central star \citep{Hollenbach-2000}. For comparison,
pre-main-sequence disc lifetimes are thought to be $5-10$~Myr \citep{Yasui-2010}.

Once the star is on the white-dwarf cooling track, any further dissipation
is set by the Poynting-Robertson drag \citep{Weidenschilling-1993},
with a time scale of the order of $10^{7}$~yr. The disc may remain
detectable for about $10^{8}$~yr, long after any sign of the planetary
nebula has disappeared.

\subsection{Resonant interactions}

\label{sub:Resonant interaction model}

\citet{Goldreich-1979} and \citet{Artymowicz-1994} describe the
resonant and nonresonant interactions between a binary system and
its CB disc using a linear perturbation theory and assuming (i) the
disc is thin ($0.01<H/R<0.1$, where $H$ and $R$ are the thickness
and the half angular momentum radius of the disc respectively) and
(ii) the nonaxisymmetric potential perturbations are small around
the average binary potential. The binary potential is expanded in
a series as 
\begin{equation}
\Phi(r,\theta,t)=\Sigma_{m,l}\phi_{m,l}(r){\rm exp}\left[i\; m\left(\theta-(l/m)\Omega_{b}t\right)\right],
\end{equation}
 of which only the real part is relevant, and $l$ and $m$ are integers.
$\Omega_{b}$ is the orbital angular frequency which, in an inertial
frame, is $\Omega_{b}=\left(\frac{G\; M}{a^{3}}\right)^{1/2}$, where
$a$ is the semi-major axis of the system, and $M$ its total mass.
The individual potential harmonics $\phi_{m,l}(r)$ rotate uniformly
with pattern speed 
\begin{equation}
\Omega_{p}=(l/m)\Omega_{b}.
\end{equation}
A resonance occurs in the disc when the pulsation $\kappa$ of radial
motion of a disc particle (on an epicyclic orbit with azimuthal pulsation
$\Omega$) is commensurate with the angular frequency $(\Omega-\Omega_{p})$
in the frame rotating with component $(l,m)$ of the perturber, 
\begin{equation}
m\,(\Omega-\Omega_{p})=\pm\kappa,
\end{equation}
where the positive/negative sign corresponds to the outer/inner Lindblad
resonance (LR) respectively. A corotation resonance (CR) occurs when
$\Omega=\Omega_{p}$. These resonant interactions result in the excitation
of density waves located at radii 
\begin{equation}
r_{{\rm CR}}=(m/l)^{2/3}a,
\end{equation}
 and 
\begin{equation}
r_{{\rm LR}}=[(m\pm1)/l]^{2/3}a.\label{Eq:LR}
\end{equation}

Many studies have investigated whether these resonances increase the
orbital eccentricity \citep{Artymowicz-1991,Artymowicz-1994,Frankowski-2007a}.
We apply the model of \citet{Lubow-1996} for small and moderate eccentricities
($e\lesssim0.2$). This model is based on results of SPH simulations
which show that for a disc in which the kinematic viscosity, $\nu$,
is independent of the density, the torque is independent of resonance
strength and width. The resonant torque can then be estimated from
the rate of change of the disc angular momentum. The viscous torques
within the disc balance the gravitational torque generated by resonances
in the inner part of the disc and redistribute mass and angular momentum
throughout the disc. The viscous torque is 
\begin{equation}
\dot{J_{\mathrm{d}}}=J_{\mathrm{d}}/\tau=M_{d}\Omega_{b}\nu,\label{eq:Torque from SPH}
\end{equation}
where the viscous timescale $\tau=R^{2}/\nu=\alpha^{-1}(H/R)^{-2}\Omega^{-1}$
is of the order of $10^{5}$~yr, and $M_{d}$, $\nu$ and $\alpha$
are the mass, kinematic viscosity and the viscosity parameter of the
disc respectively. The viscosity parameter is typically $\alpha\approx0.1$. 

\begin{figure}
\hspace{0.5cm}\includegraphics[bb=50bp 100bp 500bp 770bp,angle=-90,scale=0.35]{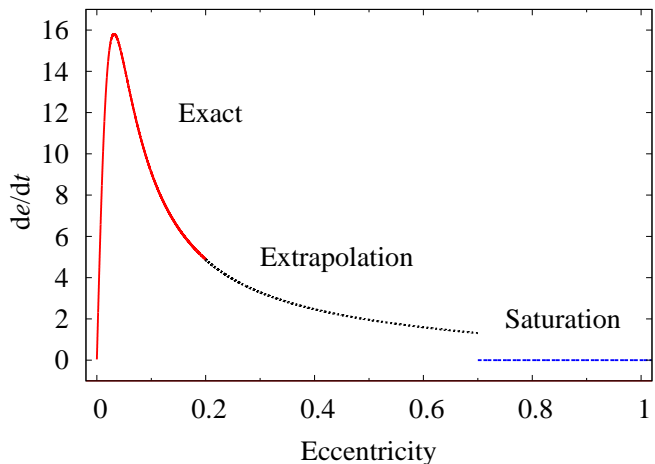}\vspace{0.5cm}

\caption{\label{fig:edot}$\mathrm{d}e/\mathrm{d}t$ in units of $\mathrm{d}\ln a/\mathrm{d}t$
as a function of $e$ (Eq.~\ref{eq:edot_combined_models}) based
on a model for very low and moderate eccentricities \citep{Lubow-1996,Lubow-2000},
for a circumbinary disc with a viscosity parameter $\alpha=0.1$.
The curve is divided in three regions. For eccentricities from 0 to
0.2, our model is strictly valid (red solid curve). For eccentricities
from 0.2 to 0.7, the model is extrapolated outside its validity range
according to an efficiency decreasing as $1/e$ (black dotted curve).
At higher eccentricities, $\mathrm{d}e/\mathrm{d}t=0$ to account
for the saturation behavior (blue dashed curve).}
\end{figure}

The variation of the orbital separation due to the resonant interaction
between the binary and the CB disc is given by \citep{Lubow-1996}
\begin{equation}
\frac{\dot{a}}{a}=2\frac{\dot{J_{d}}\Omega_{p}}{J_{\mathrm{orb}}\Omega_{b}}=-\frac{2l}{m}\frac{M_{d}}{\mu}\alpha\left(\frac{H}{R}\right)^{2}\frac{a}{R}\Omega_{b},\label{eq:adot}
\end{equation}
where $J_{\mathrm{orb}}$ is the orbital angular momentum and $\mu$
is the reduced mass of the binary system. The increase of the eccentricity
due to resonances can be given as a function of $\dot{a}/a$ and depends
on the binary eccentricity. At very small eccentricities ($e\le0.1\alpha^{1/2}$),
the resonances weakly drive the eccentricity so the inner radius of
the CB disc is maintained by the $m=l$ resonances that take place
close to the binary at $r/a\lesssim1.7$. The eccentricity pumping
in that regime increases with $e$ and is estimated \citep{Lubow-1996}
to proceed as 
\begin{equation}
\dot{e}=-\frac{50e}{\alpha}\frac{\dot{a}}{a},
\end{equation}
up to a maximum of $\dot{e}=5\alpha^{-1/2}\dot{a}/a$ at $e\simeq0.1\alpha^{1/2}$.
At larger eccentricities ($0.1\alpha^{1/2}\lesssim e\lesssim0.2$),
the $m=2,\, l=1$ resonance is the strongest contribution to the eccentricity
driving and the disc is progressively pushed away as the binary eccentricity
increases. In that regime, the eccentricity pumping decreases as $1/e$.
The eccentricity growth diminishes as $1/e$ until $e\simeq0.5-0.7$,
at which point resonances that damp the eccentricity begin to dominate
\citep[e.g.][]{Rodig-2011}. To include this effect, we assume the
$e$-pumping to be inefficient ($\dot{e}=0$) for $e\ge0.7$ (note
however that $\dot{a}$ is not necessarily 0). The $e-$pumping efficiency
for very small and moderate eccentricities (i.e. for $e\lesssim0.2$)
is shown in Fig.~\ref{fig:edot} and is given by (see \citealp{Lubow-1996,Lubow-2000})

\begin{equation}
\dot{e}=\frac{2\left(1-e^{2}\right)}{e+\frac{\alpha}{100e}}\left(\frac{l}{m}-\frac{1}{\sqrt{1-e^{2}}}\right)\frac{\dot{a}}{a}.\label{eq:edot_combined_models}
\end{equation}

The inner radius of the disc, $r\mathrm{_{in}}$, is determined by
the condition that the resonant torque, which tends to push the disc
away, overcomes the viscous torque which acts to fill the inner region.
\citet{Artymowicz-1994} estimate from SPH simulations the disc inner
radius at different binary eccentricities and for different Reynolds
number of the disc gas, ${\cal R}=(H/R)^{-2}\alpha^{-1}$ (see Fig.~3
of \citealp{Artymowicz-1994}) which we fit with 
\begin{equation}
r_{\mathrm{in}}(e,{\cal R})=1.7+\frac{3}{8}\log({\cal R}\sqrt{e})\,\mathrm{AU.}\label{eq:rin fit Arty 94}
\end{equation}

Finally, the variation of the orbital angular momentum in the centre-of-mass
frame is given by 
\begin{equation}
\dot{J}_{\mathrm{orb}}=J_{\mathrm{orb}}\frac{\dot{a}}{2a}-J_{\mathrm{orb}}\frac{e\dot{e}}{(1-e^{2})}.
\end{equation}
Our model is in reasonable (within a factor two) agreement with the
SPH results of \citet{Artymowicz-1991}, who give $\dot{a}/a\simeq-4.3\times10^{-4}\Omega_{b}M_{d}/M_{b}$
and $\dot{e}e\simeq1.9\times10^{-3}\Omega_{b}M_{d}/M_{b}$ at $e=0.1$,
for $\mu=0.3$, $\alpha=0.1$, ${\cal R}=10^{3}$, with a CB disc
extending from $r=1.7a$ to $r=6a$, with a surface density distribution
$\sigma\propto r^{-1}$. Our model gives $\dot{a}/a\simeq-2.8\times10^{-4}\Omega_{b}M_{d}/M_{b}$
and $\dot{e}e\simeq2.5\times10^{-3}\Omega_{b}M_{d}/M_{b}$.

\subsection{Eccentricity gap}

Interestingly, pre-MS and MS systems lack nearly circular orbits at
long periods (called the \emph{eccentricity gap}; see \citealp{Mathieu-1992,Mathieu-94}),
whereas short-period systems are circularised by tides. Although the
eccentricity distribution of pre-MS and MS systems is likely the result
of numerous processes such as binary formation, stellar encounters
and tides, interaction with a CB disc certainly occurs as well. Discs
commonly observed in pre-MS systems are similar to those observed
around post-AGB stars \citep{DeRuyter-2006}, with masses from 0.004
to 0.3 M$_{\odot}$. \citet{Artymowicz-1991}, \citet{Artymowicz-1994},
\citet{Mathieu-1995} and \citet{Lubow-1996} suggest that they pump
the binary eccentricity, which explains the eccentricity gap. Note
however the surprising exception of GW Orionis, a system still embedded
in a disc of $M_{d}\approx0.3$~M$_{\odot}$, with a period of 242~d
and an unexpected low eccentricity of $e\approx0.04\pm0.06$ \citep{Mathieu-1991}.
Following the evolution of pre-MS with disc-binary interaction is
however not the scope of this paper.

Note that an eccentricity gap seems also present in post-AGB binaries
(see Fig.~\ref{fig:typical-evolutionary-track-with-disc}) with periods
longer than $10^{3}\,$d, in which tides are inefficient.

\section{Modelling post-AGB binaries with circumbinary discs}

\label{sec:Modelling post-AGB with CB discs}

We introduce the CB-disc properties and apply our model of Section~\ref{sub:Resonant interaction model}
to derive the evolution of post-AGB systems in the period-eccentricity
plane.

\subsection{Circumbinary disc model}

\label{sub:Circumbinary disk Model}

\begin{figure}
\vspace{-5cm}\hspace{0.5cm}\includegraphics[bb=50bp 100bp 500bp 770bp,width=0.22\paperwidth]{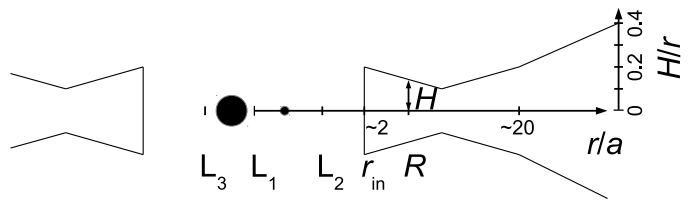}\vspace{1.5cm}

\caption{\label{fig:Disc Geometry}Disc geometry as described in \citet{Dullemond-2001}.
L$_{1}$, L$_{2}$ and L$_{3}$ are the Lagrangian points, where forces
cancel out. $r_{\mathrm{in}}$, $R$ and $H$ are the inner radius,
the half angular-momentum radius and the thickness of the disc, respectively.
The given radii are indicative.}
\end{figure}

The post-AGB disc properties can be derived from spectral-energy-distribution
(SED) modelling in the near-IR. Observed disc masses range from about
$10^{-4}$ to $10^{-2}$ M$_{\odot}$ \citep{Gielen-2007a}. The relative
thickness near the inner edge is $H/R=0.1-0.25$ \citep{Dullemond-2001}.
However, as our model is only valid for thin discs, we assume $H/R=0.1$.

The inferred location of the inner disc edge mainly depends on the
opacity of the grains which in turn is related to their chemistry
and size distribution. Metallic iron has no spectral feature but a
large opacity so, depending on the adopted abundance of iron, the
SED can be fitted with different inner radii ranging from about $2$
to $10$~AU. These values are in good agreement with our predicted
inner radius (Eq.~\ref{eq:rin fit Arty 94}) assuming typical separations
of post-AGB systems from about 0.5 to 5~AU. The outer radius ranges
from about $100$ to a few thousands AU \citep{DeRuyter-2006} and
is taken to be $r_{\mathrm{out}}=500$~AU. The surface density distribution
decreases with distance from the centre of the disc, $\sigma(r)\propto r^{\delta}$,
with $-2\lesssim\delta\lesssim-1$ \citep{DeRuyter-2006}. In our
model, we choose $\sigma(r)\propto r^{-2}$ such that the half-angular-momentum
radius is $R=\sqrt{r_{\mathrm{in}}r_{\mathrm{out}}}$.

Since when $e=0$ the pumping of the eccentricity is ineffective ($\dot{e}=0$),
we assume that perturbations of the orbit, e.g. due to the CB disc
or to stellar pulsation lead to a minimum eccentricity of $10^{-3}$.

\subsection{Post-AGB evolution in the $e-\log P$ plane with CB disc interaction}

\label{sub:Typical track in the e-logP diagram}

\begin{table}
\begin{centering}
\begin{tabular}{|c|c|c|}
\hline 
Parameter & Range & Adopted\tabularnewline
\hline 
$M{}_{\mathrm{d}}$/M$_{\odot}$ & $10^{-4}-10^{-2}$ & $10^{-2}$\tabularnewline
\hline 
$t_{\mathrm{d}}$/yr & $2.5\times10^{4}-10^{5}$ & $2.5\times10^{4}$\tabularnewline
\hline 
$\alpha$ & $0.01-0.1$ & 0.1\tabularnewline
\hline 
$H/R$ & $0.1-0.2$ & 0.1\tabularnewline
\hline 
$r_{\mathrm{in}}$ &  & Eq.~\ref{eq:rin fit Arty 94}\tabularnewline
\hline 
$r_{\mathrm{out}}$/AU & $100-2,000$ & 500\tabularnewline
\hline 
$\mu$/M$_{\odot}$ &  & 0.3\tabularnewline
\hline 
\end{tabular}
\par\end{centering}

\vspace{0.2cm}\caption{\label{tab:Models}Possible range and adopted values of the parameters,
where $M_{d}$, $t_{d}$, $\alpha$, $H/R$, $r_{\mathrm{in}}$ and
$r_{\mathrm{out}}$ are the mass, the lifetime, the viscosity parameter,
the thickness, the inner radius and the outer radius of the circumbinary
disc, respectively (see also Fig.~\ref{fig:Disc Geometry}). $\mu$
is the reduced mass of the binary. See Section~\ref{sub:Circumbinary disk Model}
for more details.}
\end{table}
\begin{figure}
\hspace{0.5cm}\includegraphics[bb=50bp 100bp 500bp 770bp,angle=-90,scale=0.35]{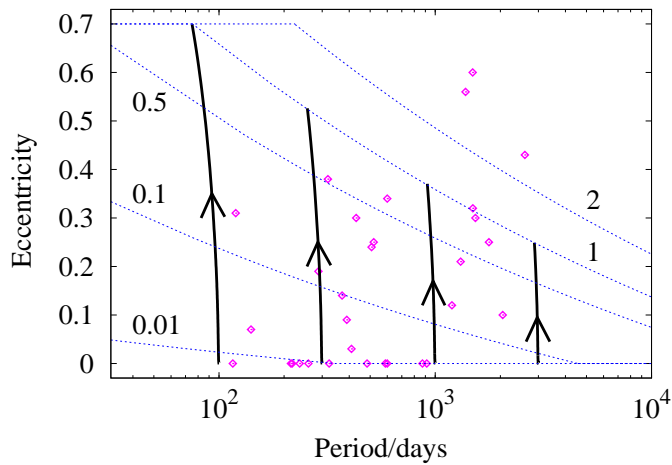}\vspace{0.5cm}

\caption{\label{fig:typical-evolutionary-track-with-disc}Evolutionary tracks
during the post-AGB phase (black lines) in the $e-\log P$ plane including
the interaction with a circumbinary disc as described by \foreignlanguage{british}{Eqs}.~\ref{eq:adot}
and \ref{eq:edot_combined_models} (see also Fig.~\ref{fig:edot})
with the adopted circumbinary-disc properties of Table~\ref{tab:Models}.
The\textcolor{red}{{} ${\color{cyan}{\color{magenta}\mathbf{\diamond}}}$}
symbols are observed post-AGB systems. The blue dashed lines show
the final eccentricity corresponding to eccentricity-pumping as given
by Eq.~\ref{eq:edot_combined_models}, modulated by the \foreignlanguage{british}{labelled}
values (1 corresponds to the adopted parameters). The timescale needed
to pump the eccentricity can be estimated from Eq.~\ref{eq:t_e-pumping}.}
\end{figure}

In Fig.~\ref{fig:typical-evolutionary-track-with-disc} we show four
evolutionary tracks in the $e-\log P$ plane of post-AGB systems with
initial periods of 100~d, 300~d, $1,000$~d and $3,000$~d, which
interact with a CB disc with properties as given in column 3 of Table~\ref{tab:Models}.
The binary systems are characterised by a reduced mass $\mu=0.3\,$M$_{\odot}$.
Their orbits are initially circular because tides are very efficient
during the AGB phase and circularise all systems with $P\lesssim2,000\,$d.
Longer-period systems can remain eccentric at the start of the post-AGB
phase. The blue dashed lines indicate the final binary eccentricity
due to eccentricity-pumping as given by Eqs.~\ref{eq:adot} and \ref{eq:edot_combined_models}
for different disc properties. Each line corresponds to pumping modulated
by the labelled value. For example, the line labelled $0.5$ corresponds
to the case where $\dot{e}$ is decreased by a factor $2$ compared
to our adopted case (labelled $1$).

The mass and lifetime of the disc are $M_{d}=10^{-2}\,$M$_{\odot}$
and $t_{d}=2.5\times10^{4}\,$yr, respectively, which correspond to
the highest mass observed and to a lower limit of the disc lifetime.
As expected from Eqs.~\ref{eq:adot} and \ref{eq:edot_combined_models},
the $e-$pumping decreases with increasing period because of the dependence
of $\dot{e}$ on the orbital period, i.e. $\dot{e}\propto a\Omega_{b}\propto P^{-1/3}$.
However, this trend is not observed among the post-AGB stars (see
discussion in Section~\ref{sec:discussion}). Nevertheless, our simulations
can reproduce the observed eccentricities of all post-AGB systems,
except for the three systems with $P>10^{3}\,$d and $e>0.4$. Moreover,
all the systems are not equally evolved on the post-AGB phase, so
some systems may have interacted longer with their CB disc leading
to a more eccentric orbit. When extending the disc lifetime to $10^{5}\,$yr,
its expected maximum lifetime, our model predicts systems with $e=0.7$
at $P\sim10^{3}\,$d (see Eq.~\ref{eq:t_e-pumping}).

\section{Discussion}

\label{sec:Discussion}

\subsection{Model uncertainties}

The time to pump the binary eccentricity from 0 to $e_{f}$ is given
by

{\small 
\begin{align}
t(e_{f}) & =2\times10^{5}e_{f}^{2}\nonumber \\
 & \frac{\mu}{0.3}\frac{10^{-2}\mathrm{M}_{\odot}}{M_{\mathrm{d}}}\frac{0.1}{\alpha}\left(\frac{0.1}{H/R}\right)^{2}\left(\frac{P}{1000\,\mathrm{d}}\right)^{2/3}\sqrt{\frac{r_{\mathrm{out}}}{500\,\mathrm{AU}}}\,\mathrm{yr}.\label{eq:t_e-pumping}
\end{align}
}Regarding the possible range of the disc parameters (i.e. $M_{\mathrm{d}}$,
$t_{d}$, $\alpha$, $H/R$ and $r_{\mathrm{out}}$) given in Table~\ref{tab:Models},
the eccentricity-pumping efficiency can be increased by a factor 36
or decreased by a factor 2000 compared to the adopted values. Fig.~\ref{fig:typical-evolutionary-track-with-disc}
shows that for an efficency decreased by a factor more than $100$,
the CB disc has a no significant effect on the binary eccentricity
any more.

\subsection{Discussion}

\label{sec:discussion}Our model is able to reproduce most orbits
of post-AGB binaries except for the three systems at $P>10^{3}\,$d
and $e>0.4$ (Fig.~\ref{fig:typical-evolutionary-track-with-disc}).
Extending the disc lifetime to $10^{5}\,$yr, the expected upper limit,
gives enough time for the CB disc to increase the system eccentricity
to the observed value. However, systems with periods longer than about
$2,000\,$d are not efficiently circularised during the AGB phase,
so they can start the post-AGB phase in an eccentric orbit. Moreover,
as the CB disc is formed during the AGB phase, eccentricity pumping
takes place before the post-AGB phase. This leads to an underestimation
of the binary eccentricity at $P\gtrsim2,000\,$d. At shorter periods,
binary systems enter a CE which circularises the orbit before the
system starts its evolution on the post-AGB phase.

There is an important discrepancy between the expected and observed
distributions of post-AGB systems in the period-eccentricity plane.
Our model predicts systems with eccentricities that decreases with
increasing periods while the observed distribution shows the opposite.
As discussed before, this discrepancy may partly be explained by an
underestimation in our model of the eccentricity at long orbital periods.
A hypothesis to explain this discrepancy is that the properties of
the disc and the binary are correlated in such a way that $e$-pumping
is more efficient in longer-period systems. It would be a surprise
if discs that form through wind mass-transfer or remaining from CE
ejection are similar. However, our understanding of disc formation
is poor, so no \emph{a priori} correlation can be suggested. None
of the observed disc properties are correlated with the binary properties.
Nonetheless, it is surprising that no post-AGB stars with a disc are
observed in systems with orbital periods shorter than $100$~d. From
binary evolutionary models, such systems are predicted to exist as
post-CE systems and are expected to have periods from a few hours
to hundreds of days. As none are observed, CB discs may be very short-lived
or unstable at periods shorter than about $100$~d.

In our model we do not consider accretion of matter onto the central
stars or outflow from the disc. While the effects of such re-accretion
are observed in some post-AGB stars \citep{Waters-92}, our model
disc mass is assumed constant for simplicity. However, as discussed
in Section~\ref{sub:Evolution}, the post-AGB systems are observed
at least $10^{4}\,$yr after they left the AGB phase. The inferred
disc masses thus represent a lower limit of their initial masses.
Note that the most important effect of re-accretion is to slow down
the evolution along the post-AGB phase. It then takes longer for the
central star to become hot enough to efficiently evaporate the disc
gas due to ultraviolet radiation compared to single star evolution
where no accretion takes place.

The conclusions reached in this paper not only concern post-AGB systems,
but a variety of classes of post-mass transfer systems such as barium
stars, S stars, subgiant CH and CEMP binaries, which we observe long
after mass transfer finished. If CB discs play an important role in
the evolution of post-AGB systems they are also relevant for these
classes of stars. Their similar periods and eccentricities to post-AGB
stars support an identical eccentricity-pumping mechanism. We are
currently working on population synthesis models including the formation
and interaction with a CB disc, which attempts at explaining the orbital
properties of all these systems (Dermine et al. 2012, in preparation).

\section{Conclusions}

\label{sec:Conclusions}Our circumbinary-disc model describes the
resonant interaction between a disc and its central binary with which
we account for the large eccentricities observed among post-AGB stars.
Stable dust discs are detected in a large fraction of post-AGB stars
and are known to be closely related to binarity. Circumbinary discs
strongly alter the evolution of the binary system because of resonant
interactions and reaccretion from the circumbinary disc. Resonant
interactions with a circumbinary disc transfer angular momentum from
the binary orbit to the disc and increases the binary eccentricity.
We use the model described by \citet{Lubow-1996} with the disc properties
derived from the observed SED of post-AGB stars. Reaccretion of gas
deficient in refractory elements, because they condense in grains
maintained in the disc by radiation pressure, is responsible for the
depletion patterns observed at the surface of numerous post-AGB systems
and slows the evolution of the post-AGB star. 

We estimate binary post-AGB lifetimes to range from $2.5\times10^{4}\,$yr
to $10^{5}\,$yr which are long compared to single-star models that
predict only $10^{3}$ to $10^{4}\,$yr. Altough the disc lifetime
represents the major uncertainty, our model reproduces the eccentricities
of most post-AGB systems on the derived lower limit of the disc lifetime.
Three long-period systems show higher eccentricities than expected
and require more efficient eccentricity-pumping. However, extending
the disc lifetime to its expected maximum lifetime of $10^{5}\,$yr
solves the problem. On the other hand, our model underestimates the
eccentricities of long-period post-AGB stars because we assume all
systems to have initially circular orbits while systems with periods
longer than $2,000\,$d are not efficiently circularised by tides
during the AGB phase. Moreover, the circumbinary disc is formed during
the AGB phase so that eccentricity-pumping already operates on the
AGB.

Finally, Ba, S, CH and CEMP stars experience a similar evolution than
post-AGB stars and show similar period and eccentricity distributions.
This suggests that interaction with a circumbinary disc is also a
key mechanism in the evolution of Ba, S, CH and CEMP stars.
\begin{acknowledgements}
TD would like to thank Thomas Masseron, Sophie Van Eck, Philipp Podsiadlowski,
Clio Gielen, Nadya Gorlova and Steve Lubow for many useful discussions.
This work has been partially funded by an \emph{Action de Recherche
Concert\'ee }from the \emph{Direction g\'en\'erale de l'enseignement
non obligatoire et de la Recherche Scientifique -- Direction de la
Recherche Scientifique -- Communaut\'e Fran{\c c}aise de Belgique.}
\end{acknowledgements}
\bibliographystyle{aa}
\bibliography{/home/dieu/svn/dermine/Papers/references,/users/dermine/svn/dermine/Papers/references}

\end{document}